\documentstyle[preprint,prl,aps,epsfig]{revtex}
\begin{document}
\preprint{LBL-PUB-43907, July, 1999} 

\title{Interference in Exclusive Vector Meson Production in Heavy Ion
Collisions}

\author{Spencer R. Klein and Joakim Nystrand} 
\address{Lawrence Berkeley National Laboratory, Berkeley, CA 94720} 

\break 
\maketitle
\vskip -.2 in
\begin{abstract}
\vskip -.2 in 

Photons emitted from the electromagnetic fields of relativistic heavy
ions can fluctuate into quark anti-quark pairs and scatter from a
target nucleus, emerging as vector mesons.  These coherent
interactions are identifiable by final states consisting of the two
nuclei and a vector meson with a small transverse momentum.  The
emitters and targets can switch roles, and the two possibilities are
indistinguishable, so interference may occur.  Vector mesons are
negative parity so the amplitudes have opposite signs. When the meson
transverse wavelength is larger than the impact parameter, the
interference is large and destructive.

The short-lived vector mesons decay before amplitudes from the two
sources can overlap, and so cannot interfere directly. However, the
decay products are emitted in an entangled state, and the interference
depends on observing the complete final state.  The non-local wave
function is an example of the Einstein-Podolsky-Rosen paradox.

\end{abstract}
\narrowtext

In relativistic heavy ion collisions, vector mesons are copiously
produced via photon-Pomeron fusion\cite{usPRC}.  The photon and
Pomeron both couple coherently to their emitting nuclei, giving these
reactions a distinctive signature, with the final state consisting of
the two nuclei, a vector meson with low perpendicular momentum
($p_\perp$) and nothing else.  The latter requirement restricts these
interactions to peripheral collisions, usually with impact parameter
$b>2R_A$, $R_A$ being the nuclear radius.  At heavy ion colliders,
vector mesons with masses up to about $2\gamma\hbar c/R_A$, where
$\gamma$ is the Lorentz boost of each beam are produced.

The electromagnetic field has a long range, while the nuclear
(Pomeron) field has a short range compared to the size of the nucleus.
So, vector meson production occurs in the region occupied by the
Pomeron emitting ('target') nucleus. Since the production is coherent
over the entire target nucleus, it is a fairly good approximation to
treat the meson production regions as two point sources, one at the
center of each nucleus.  The situation is similar to that in a
two-source interferometer, albeit with unstable particles.  A
parity inversion switches the emitter and target. Because of
the vector meson negative parity, the two emission amplitudes have
opposite sign.  For perpendicular momentum $p_\perp \ll 1/b$, $b$
being the impact parameter, the interference is destructive.  We
calculate here the magnitude of the interference and discuss the
implications of the short vector meson lifetime, which causes the
mesons to decay before wave functions from the two sources can
overlap.

The cross sections for meson production can be calculated by
convoluting the Weizs\"acker-Williams virtual photon spectrum with the
photonuclear interaction cross section.  The photonuclear cross
section is determined from data on $\gamma p$ interactions\cite{data},
using a Glauber formalism to find the cross section for $\gamma
A\rightarrow VA$, where $A$ is the target nucleus, and V a vector
meson.  In this approach, the photons are considered to fluctuate to
virtual vector mesons, which then elastically scatter from the target
nucleus, emerging as real mesons.

Earlier calculations of production cross sections summed the cross
sections from the two nuclei and found very large cross sections
\cite{usPRC}.  Exclusive $\rho^0$ production was about 10\% of the
hadronic cross section for gold on gold collisions at a per nucleon
center of mass energy $\sqrt{S_{NN}}=200$ GeV as will be studied at
the Relativistic Heavy Ion Collider (RHIC), rising to 30\% for the
$\sqrt{S_{NN}}=5.5$ TeV lead on lead collisions at the Large Hadron
Collider (LHC).  The $\omega$ and $\phi$ production are about an order
of magnitude smaller, with $J/\psi$ cross sections of 0.3 mb (gold at
RHIC) up to 32 mb (Pb at LHC). These cross sections correspond to
production of hundreds of $\rho^0$ per second at RHIC, and hundreds of
thousand of $\rho^0$ per second with calcium beams at the LHC.  The
$J/\psi$ rates range from 0.06 Hz with gold at RHIC up to 780 Hz with
calcium at LHC. These rates are comparable to those found at current
and future meson factories.

Because of the small $b$, for low $p_\perp$ final states it is
impossible to tell which nucleus emitted the photon, and which
elastically scattered the meson.  So, the amplitudes are combined.
Because vector mesons are negative parity, the two amplitudes
subtract, rather than add, leading to destructive interference.  In
different terms, for $p_\perp b < 1$, a system of two identical
nuclei has zero dipole moment, so radiation of vector particles is
impossible.  This situation is analogous to bremsstrahlung by
identical particles\cite{nature}, where there can be almost complete
destructive interference.  However, in contrast to bremsstrahlung
photons, vector mesons are short lived, and they don't live long
enough to travel the distance $b$.  So, the vector mesons generally
decay before their wave functions can overlap. However, their decay
products can interfere.

The amplitude for vector meson production is
$A(y,p_\perp,b)e^{i\phi(y)}$, with $A(y,p_\perp,b)$ the magnitude and
$\phi(y)$ the phase at rapidity $y$. We assume that
$A(y,p_\perp,b)e^{i\phi(y)}$ is symmetric to rotations around the beam
direction.  The photon energy and perpendicular momentum are $k$ and
$k_\perp$, and the final meson momentum $p$.  All values are in the
center of mass frame, which corresponds to the laboratory frame for a
heavy ion collider.  To eliminate the directional ambiguity, we adopt
the convention that, for $y>0$, the photon energy is higher than the
Pomeron energy.  Then, $A(y>0) < A(y<0)$ and $y=\ln{(2k/M_V)}$.  These
amplitudes are determined from the cross sections calculated in
Ref.~\cite{usPRC}, using $A(y,p_\perp,b)=\sqrt{\sigma_{A +
A\rightarrow A + A + V}(y,p_\perp,b)}$, for a single production
direction.  

The final state perpendicular momentum is the sum of the photon and
Pomeron perpendicular momentum; the spectrum is a convolution of
the two sources.  The photon perpendicular momentum can be found with
the equivalent photon approximation\cite{vidovic}:
\begin{equation}
{d^3N_\gamma(k,k_\perp) \over d^2k_\perp dk } = 
{\alpha^2 Z^2 F^2(k_\perp^2 + k^2/\gamma^2) k_\perp^2 
\over 
\pi^2 (k_\perp^2 + k^2/\gamma^2)^2 }.
\end{equation}
The nuclear form factor,
\begin{equation}
F(q) = {4\pi\rho_0 \over Aq^3} \bigg[\sin(qR_A) - qR_A\cos(qR_A)\bigg]
\bigg[{1\over 1+a^2q^2}\bigg],
\label{eqff}
\end{equation}
the convolution of a hard sphere and a Yukawa potential with range
$a=0.7$ fm , is an excellent fit to a Woods-Saxon density
distribution\cite{usPRC}. Here, $\rho_0$ is the nuclear density and
$A$ the atomic number.  For a given $k$, $k_\perp$ is independent of
$b$, with $d^3N_\gamma(k,k_\perp)/d^2k_\perp dk$ rising from 0 at
$k_\perp=0$ to a maximum at $k_\perp=k/\gamma$ and then dropping as
$k_\perp$ rises further.

Since the Pomeron range is short, its perpendicular momentum spectrum
is determined by Eq. (\ref{eqff}).  Figure 1 shows the photon and
Pomeron perpendicular momentum spectra, along with their convolution,
for $y=0$, and $y=-2$.  Diffractive fringes are visible in the Pomeron
spectrum; since the photon contribution to $p_\perp$ is usually small,
these fringes also appear in the final $p_\perp$ spectrum.

The production phase depends on the process.  Since the soft Pomeron
represents the absorptive part of the cross section, the amplitude for
photon-Pomeron fusion should be largely imaginary, with a small real
part because the cross section rises slowly with photon
energy\cite{leith}. Parameterizing $\sigma(\gamma N\rightarrow VN)\sim
W^{\epsilon}$, where $W$ is the $\gamma$-nucleon center of mass
energy, $\phi(y)=\tan^{-1}(\pi\epsilon/4)$, independent of
$y$\cite{forshaw}. For the light mesons, $\epsilon=0.22$ so $\phi\sim
10^0$.  For the $J/\psi$, $\sigma(\gamma N\rightarrow VN) \sim
W^{0.8}$, inconsistent with the soft Pomeron; the steep rise may be
due to threshold behavior or signal the breakdown of the soft Pomeron
model\cite{nicolo}; in either case, the $J/\psi$ phase angle must be
treated with caution.

For the $\rho^0$ and $\omega$, photon-meson (primarily the $f_0$
\cite{leith}) fusion also contributes to the cross section, which may
be parameterized as $XW^\epsilon + YW^{-\eta}$, where $\eta > 1$, so
the photon-meson contribution decreases as $W$ rises.  Besides
increasing the cross section, this causes $\phi(y)$ to vary with $y$.
Because the phase has only been measured as an average over photon
energy\cite{review}, we will treat $\phi(y)$ as a constant.  Of
course, at $y=0$, the energies are equal; as $|y|$ rises, the phase
difference could become significant.

For two nuclei at points $\vec{x_1}$ and $\vec{x_2}$, the amplitude
$A_0$ for observing a vector meson at a distant point $x_0$ is found
by approximating the vector mesons by plane waves.  The meson momentum
$\vec{p}$ is determined by $p_\perp$ and $k$ via $p_{||}=k -
m_V^2/4k$.
\begin{equation}
A_0(x_0,\vec{p},b) = 
A(p_\perp,y,b)e^{i[\phi(y)+ \vec{p}\cdot (\vec{x_1}-\vec{x_0})]} -
A(p_\perp,-y,b)e^{i[\phi(-y)+ \vec{p}\cdot (\vec{x_2}-\vec{x_0})]}
\label{ampone}
\end{equation}
with the negative sign because of the negative parity.  Because of
their short lifetimes, the vector mesons will decay before reaching
$x_0$.  In fact, they will generally decay before their wave functions
can overlap.  However, their decay products can interfere; with
$\vec{p}$ now being the sum of the product momenta.  The resulting
virtual interference pattern depends only on $\vec{p}\cdot\vec{b}$.
The decay is incorporated by multiplying $A(p_\perp,y,b)$ by the decay
amplitude, including the branching ratios, lifetime and angular
dependence.  Since the angular distributions are the same for the two
production directions, the decay amplitudes are the same for both
directions, and does not affect the observed cross sections.

Since the decay products fly off in different directions at
relativistic velocities, the production amplitudes from the two
sources can only interfere if the final state wave function is
nonlocal\cite{pra}.  For example, in $J/\psi\rightarrow e^+e^-$, the
electron and positron are nearly back-to-back, and the $e^+$ (or
$e^-$) amplitudes from the two sources cannot overlap each other until
they are a good distance from their parent $J/\psi$. By the time the
$e^+$ (and $e^-$) from the two sources overlap, the $e^+$ and $e^-$
are well separated, and any interference involving both the $e^+$ and
$e^-$ requires a non-local wave function.

The final state wave function is entangled, with the form
$\exp(ik_1\cdot x_1)\exp(ik_2\cdot x_1) - \exp(ik_2\cdot
x_2)\exp(ik_2\cdot x_2)$ where $k_1$ and $k_2$ are the individual
decay product momenta (for a 2 body decay), and cannot be factored
into individual particle wave functions.  This interference is visible
by observing the complete final state, but not by examining the
individual decay products.  This non-locality is an example of the
Einstein-Podolsky-Rosen paradox\cite{EPR}.  However, the correlation
depends on $p_\perp$, which is a continuous variable, rather than a
discrete variable like spin, as is commonly considered.

The individual decay products have much higher momenta than their
parent meson, and in principle could be used to reconstruct the meson
decay point, and hence determine which nucleus emitted the meson.
However, the position measurement would obscure their momentum enough
to eliminate the interference pattern, as occurs with a
two-slit interferometer\cite{prb}.

In the center of mass frame, production is nearly simultaneous
(within $\Delta t = b/c\gamma$), so
time drops out.  Since $A$ is defined to be real,
\begin{equation}
\sigma(p_\perp,y,b) = A^2(p_\perp,y,b) +
A^2(p_\perp,-y,b) - 2A(p_\perp,y,b)A(p_\perp,-y,b) 
\cos[\phi(y)-\phi(-y) + \vec{p}\cdot \vec{b}]
\label{eqaa}
\end{equation}
At midrapidity, $y=0$, the two source contributions are equal and the
observed cross section is
\begin{equation}
\sigma(p_\perp,y=0,b)= 2A^2(p_\perp,y=0,b) (1-\cos[\vec{p}\cdot
\vec{b}]).
\end{equation}
For a given $b$, $\sigma$ oscillates with period $\Delta p_\perp = \hbar/b$.
When $\vec{p}\cdot\vec{b}\ll 1$, the interference is destructive, and
there is little emission.  

Since $b$ is in principle measurable by examining the outgoing ions,
it is an observable, and we integrate the cross section (not
amplitude) over all possible $b$ to get the total production.  Figure
2 shows $d^2N/dp_\perp^2$ at $y=0$ for gold and silicon beams at RHIC
and calcium at LHC, with and without interference.

In theory, one could choose to measure the perpendicular momentum
transfer from the two nuclei instead of $b$. Then, for roughly
$p_\perp > 10$ MeV/c, the larger momentum transfer will
usually come from the target.  This determination will break the
symmetry, reducing the magnitude of the interference. Of course, near
$p_\perp=0$, the photon and Pomeron perpendicular momenta have very
similar magnitudes, so the interference will remain.  Because the
initial nuclear momenta are not well known, this possibility
seems at best theoretical.

With gold beams at RHIC, vector meson production occurs with median
impact parameters for $\rho^0$, $\omega$, $\phi$ and $J/\psi$
production of 46 fm, 46 fm, 38 fm and 23 fm respectively,
corresponding to $p_\perp<4$ MeV/c for the light mesons and
$p_\perp<10$ MeV/c for the $J/\psi$.  Below these $p_\perp$ values,
interference is large.  With lighter nuclei, the average impact
parameters are slightly smaller, leading to higher momentum cutoffs.
At the LHC, impact parameters are much larger, 170-290 fm for light
vector mesons and 44-68 fm for the $J/\psi$, and the reduction is
large only for quite small values of $p_\perp$.

Despite the dramatically different $p_\perp$ spectrum, the the overall
rate at $y=0$ is unaffected by the interference.  This is because
$\langle p_\perp\rangle > \langle \hbar/b \rangle$, so the
$\cos[\vec{p}\cdot\vec{b}]$ oscillations average out, leaving the rate
almost unaffected.  Since the $p_\perp$ spectrum is almost independent
of $y$, this remains true at other rapidities, so the total cross
section is also unchanged.

Figure 3 shows $d^2N/dp_\perp^2$ at $y=1$ for the same three cases.
The magnitude of the interference is reduced because of the different
amplitudes for the two directions.  Besides the overall amplitudes,
the spectrum is slightly different, because the photons have different
energies, and so the $p_\perp$ spectra are slightly different.

This interference can be studied with the decays
$\rho^0\rightarrow\pi^+\pi^-$, $\omega\rightarrow\pi^+\pi^-$,
$\phi\rightarrow K^+K^-$ and $J/\psi\rightarrow e^+e^-$ which are
readily reconstructible.  Because of the high rates, the backgrounds
to these processes should be small; simulations indicate that this is
indeed the case, at least for a large acceptance
detector\cite{usRHIC}.

There are a few details that may slightly reduce the interference.
Because the photon emission is electromagnetic, it is possible for the
emitting nucleus to be excited into a giant dipole resonance (GDR);
since the Pomeron is uncharged, excitation is unlikely for the
scattering nucleus.  However, even for the photon emitting nucleus,
the excitation probability is relatively low.  Excitation is much more
likely if the nuclei exchange an additional photon between themselves,
with equal excitation probabilities for both nuclei, and not affecting
the interference\cite{baurrev}.  Radiative corrections and other
higher order processes could also reduce the interference. These
factors should be small because the nuclear trajectories are barely
affected by the interaction.  Likewise, backgrounds from non-resonant
(but coherent) photonuclear and two-photon processes should be small.

However, these backgrounds may introduce additional interference terms
which may complicate the picture.  For example, the process
$\gamma\gamma\rightarrow e^+e^-$ can interfere with $e^+e^-$ from
$J/\psi$ decays; the rate for the $\gamma\gamma\rightarrow e^+e^-$ 
to mimic $J/\psi$ decays is small, but perhaps visible.

In conclusion, interference between emission between two colliding
nuclei dramatically changes the perpendicular momentum spectrum of
exclusively produced vector mesons, suppressing production of vector
mesons at low $p_\perp$. Because the vector mesons decay before their
wave functions can overlap, the interference involves all of the decay
products, which never overlap with each other. So, the
decay products have an entangled, non-local wave
function.

We would like to acknowledge useful conversations with Stan Brodsky
and J\o rgen Randrup.  This work was supported by the US DOE, under
contract DE-AC-03-76SF00098.

\begin{figure}
\label{fpperp}
\setlength{\epsfxsize=0.75\textwidth}
\centerline{\epsffile{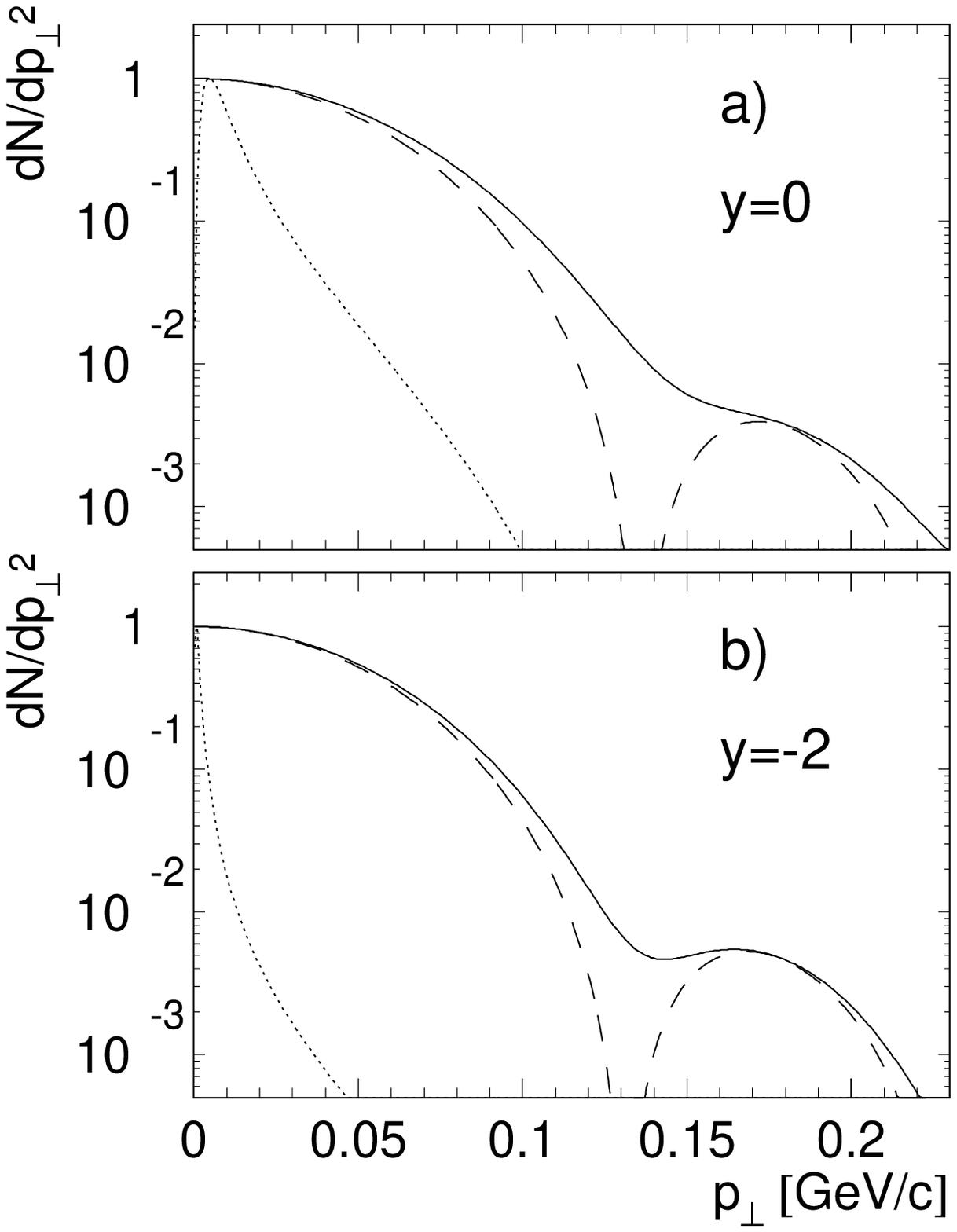}}
\caption[]{Perpendicular momentum spectra for photons (dotted curves),
Pomerons (dashed curves) and the final state vector mesons (solid
curves) at (a) $y=0$, and (b) $y=-2$ (corresponding to $k=69$ MeV in
the lab frame) for $\phi$ production in gold collisions at RHIC.
The curves are each normalized to a maximum $dN/dp_\perp^2=1$.
Clear diffraction minima appear in the Pomeron spectra.  Since
$k_\perp$ is small compared to $p_\perp$, the minima remain
visible in the $\phi$ $p_\perp$ spectrum.}
\end{figure}

\begin{figure}
\label{fphotonflux}
\setlength{\epsfxsize=1.0\textwidth}
\centerline{\epsffile{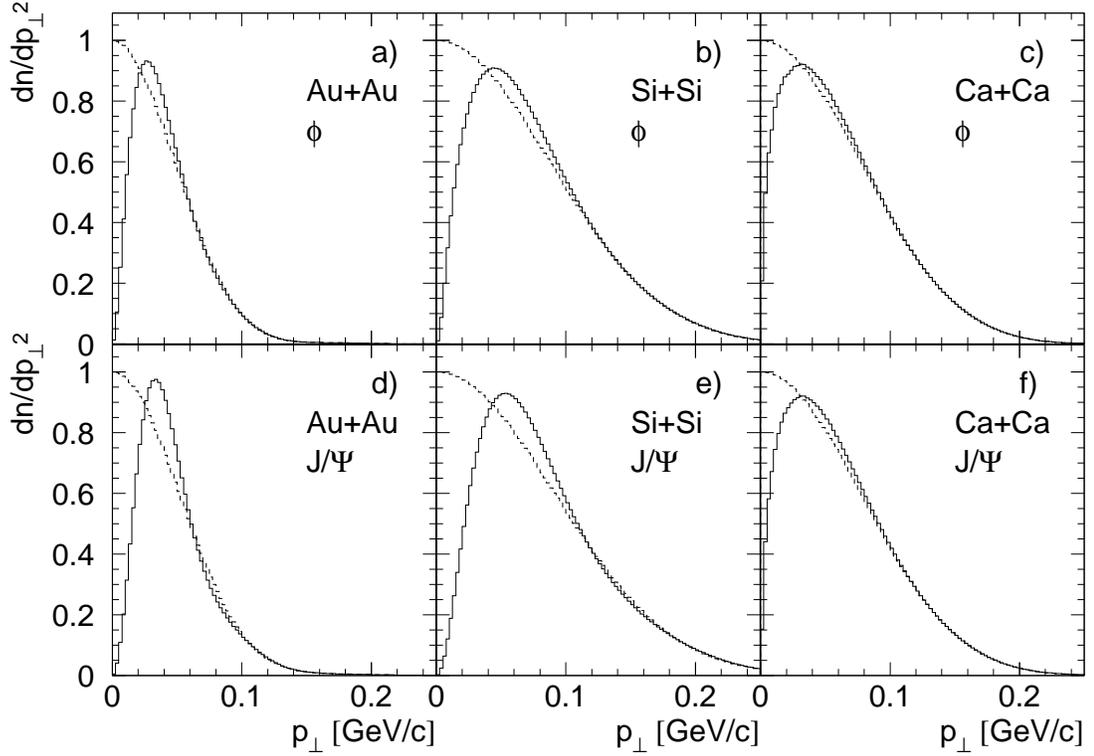}}
\caption[]{Expected $p_\perp$ spectra for reconstructed $\phi$ and
$J/\psi$ mesons at y=0 with (a,d) gold beams at RHIC, (b,e) silicon
beams at RHIC and (c,f) calcium beams at LHC.  The solid histograms
include interference, while the dotted ones do not.  Because of the
smaller impact parameters production is peaked at higher $p_\perp$ in
(b,e) than in (a,d).  Because of the smaller impact parameters in
(b,e), the interference dip extends to higher $p_\perp$ than in (a,d).
In (c,f), the energies are higher, leading to higher impact
parameters, pushing the interference dip to lower $p_\perp$.  The
figure is normalized so that, without interference $dn/dp_\perp^2=1$
at $p_\perp=0$; the rates are given in Ref. \cite{usPRC}.}
\end{figure}

\begin{figure}
\label{fphotonflux1}
\setlength{\epsfxsize=1.0\textwidth}
\centerline{\epsffile{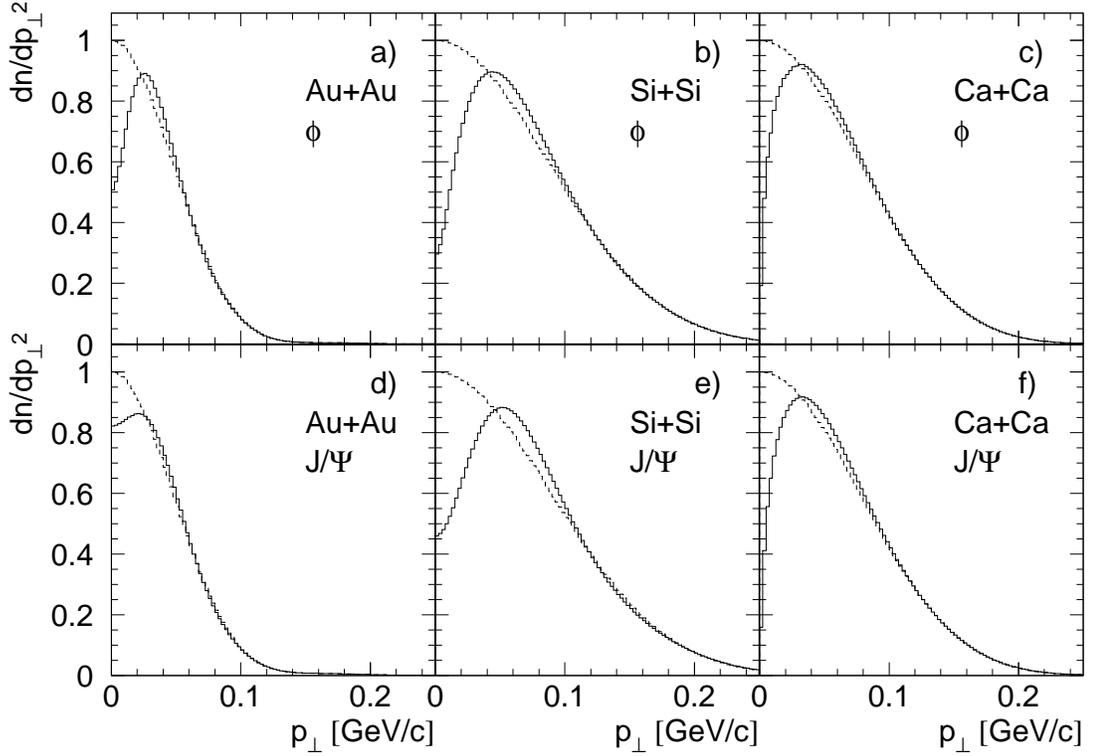}}
\caption[]{Expected $p_\perp$ spectra for reconstructed $\phi$ and
$J/\psi$ mesons at y=1 with (a,d) gold beams at RHIC, (b,e) silicon
beams at RHIC and (c,f) calcium beams at LHC.  The solid histograms
include interference, while the dotted ones do not.  Because of the
difference in amplitudes, $dN/dp_\perp^2\ne 0$ at $p_\perp=0$.  The
size of the dip at $p_\perp=0$ depends on the ratio of amplitudes at
$y=1$ and $y=-1$; it is larger for lighter nuclei and higher collision
energies.  This figure is normalized so that, without interference
$dn/dp_\perp^2 =1$ at $p_\perp=0$; the rates are given in
Ref.~\cite{usPRC}.}
\end{figure}

\end{document}